# Micromagnetic Monte Carlo method with variable magnetization length based on the Landau-Lifshitz-Bloch equation for computation of large-scale thermodynamic equilibrium states

Serban Lepadatu[*]

*Jeremiah Horrocks Institute for Mathematics, Physics and Astronomy, University of Central Lancashire, Preston PR1 2HE, U.K.*

**Abstract**

An efficient method for computing thermodynamic equilibrium states at the micromagnetic length scale is introduced, using the Markov chain Monte Carlo method. Trial moves include not only rotations of vectors, but also a change in their magnetization length. The method is parameterized using the longitudinal susceptibility, reproduces the same Maxwell-Boltzmann distribution as the stochastic Landau-Lifshitz-Bloch equation, and is applicable both below and above the Curie temperature. The algorithm is fully parallel, can be executed on graphical processing units, and efficiently includes the long range dipolar interaction. This method is generally useful for computing finite-temperature relaxation states both for uniform and non-uniform temperature profiles, and can be considered as complementary to zero-temperature micromagnetic energy minimization solvers, with comparable computation time. Compared to the dynamic approach it is shown the micromagnetic Monte Carlo method is up to almost 20 times faster. Moreover, unlike quasi-zero temperature approaches which do not take into account the magnetization length distribution and stochasticity, the method is better suited for structures with unbroken symmetry around the applied field axis, granular films, and at higher temperatures and fields. In particular, applications to finite-temperature hysteresis loop modelling, chiral magnetic thin films, granular magnetic media, and artificial spin ices are discussed.

[*] SLepadatu@uclan.ac.uk

# I. Introduction

Monte Carlo computations are widely used in many fields of research, including statistical physics and atomistic spin lattice modelling. For the Ising and Heisenberg Hamiltonian spin lattice models, the Metropolis Monte Carlo [1] algorithm has proved popular. This allows computation not only of magnetic parameters temperature dependences and phase transition temperatures, but also hysteresis loops [2]. In micromagnetics modelling a common need is computation of relaxed magnetization states. For zero-temperature models this may be accomplished efficiently using energy minimization solvers, such as a steepest descent method [3]. Finite-temperature micromagnetic models include the stochastic Landau-Lifshitz-Gilbert equation (sLLG) [4], as well as the Landau-Lifshitz-Bloch (LLB) equation [5] and its stochastic forms (sLLB) [6,7]. Whilst the sLLG equation is a reasonable approximation at low temperatures, it fails to take into account the magnetization length distribution, which has been shown to play an important role in magnetization reversal [8], including linear and elliptical reversal mechanisms [9], particularly for temperatures approaching the Curie point. Thus, whilst the sLLB equation may be used to compute finite-temperature relaxation states, this dynamic approach to relaxation is very inefficient and requires lengthy computation times. Alternative methods, such as a Monte Carlo approach, are required to efficiently compute relaxation states of thermodynamic equilibrium, which should ideally be applicable across the entire temperature range, both below and above the Curie temperature; for this reason correct reproduction of the magnetization length probability distribution at the micromagnetic length scale is essential.

A previous work extended the Monte Carlo method to the micromagnetic length scale, however the magnetization length distribution was not taken into account [10]; the resulting thermodynamic equilibrium properties are the same as those produced by the sLLG equation. Other works along this line also include a micromagnetic one-dimensional model [11], as well as a two-dimensional model [12]. Alternatively, a thermodynamic rescaling for the atomistic cluster Monte Carlo approach was also discussed [13]. A Monte Carlo method was also used for the Landau-Heisenberg Hamiltonian, applied to the study of phase transitions in FeRh including magnetic and non-magnetic contributions [14]. A micromagnetic hybrid Monte Carlo method was introduced previously [15], based on the hybrid (molecular dynamics/Langevin) Monte Carlo approach [16], which reproduces a Boltzmann distribution for the free energy. At the micromagnetic length scale however, particularly when approaching the Curie temperature, the distribution of the free energy does not follow a Boltzmann distribution, but a Maxwell-Boltzmann type distribution [7]. Moreover, for the purposes of computing thermodynamic equilibrium states at the micromagnetic length scale the hybrid Monte Carlo approach is inefficient, since an ensemble of conjugate momenta must be evolved over many iterations (typically 100 or more), before a new micromagnetic magnetization configuration may be accepted or rejected.

Here we describe a micromagnetic Monte Carlo (MMC) method based on the Markov Chain approach, thus similar to the Metropolis Monte Carlo method [1] commonly used for atomistic spin lattice simulations, but including trials moves for magnetization length change. The resulting method not only correctly reproduces the expected Maxwell-Boltzmann distribution for the free energy, but generates new micromagnetic magnetization configurations every iteration. The algorithm has been implemented in Boris [17], and is publically available and open-source [18]. In Section II the algorithm and related theory are described and tested. In Section III the inclusion of the long range demagnetizing interaction in the parallel MMC implementation is described, and tested by computation of hysteresis loops and comparison with established methods. In Section IV the application of MMC to chiral magnetic films with Dzyaloshinsky-Moriya interaction (DMI) [19,20] is discussed. Finally, in Section V the application of MMC to the study of artificial spin ices is addressed, before summarising possible future developments in the concluding remarks.

## II. Micromagnetic Markov Chain Monte Carlo Method

For atomistic spin lattice models, where the spins have fixed length, at thermodynamic equilibrium the internal energy $E$ follows a Boltzmann distribution, $\exp(-E/k_B T)/Z$, where $Z$ is the partition function, $k_B$ is the Boltzmann constant, and $T$ is the temperature. At the micromagnetic length scale however, where the magnetization length can vary due to thermodynamic averaging over many atomistic spins, the distribution of the free energy follows a Maxwell-Boltzmann type distribution [7], given in Equation (1).

$$f(\mathbf{m}) = m^2 \exp(-F(\mathbf{m})/k_B T)/Z,$$
$$Z = \sum_i m_i^2 \exp(-F(\mathbf{m}_i)/k_B T). \tag{1}$$

Here $\mathbf{m} = \mathbf{M}/M_{S0}$, where $\mathbf{M}$ is the magnetization vector and $M_{S0}$ is the zero-temperature saturation magnetization, and $m = |\mathbf{m}|$. $F(\mathbf{m})$ is the micromagnetic free energy, and contains a number of interactions, including applied field, exchange interaction, DMI, magnetocrystalline anisotropy, and demagnetizing interaction. The distribution in Equation (1) is reproduced by the sLLB equation [7], as well as atomistic spin lattice simulations where the magnetization is computed by thermodynamic averaging of atomistic spins, as we have verified numerically. Whilst the sLLB equation may be used to compute relaxed magnetization states in thermodynamic equilibrium, as required for example for finite-temperature hysteresis loop modelling, this approach is very inefficient, particularly for large-scale simulations. A far more efficient approach may be obtained using a Monte Carlo method. In particular, we wish to establish a method which generates the distribution in Equation (1), using a similar method to the Metropolis Monte Carlo [1] employed for atomistic spin lattices, by generating new micromagnetic magnetization configurations every iteration. For atomistic simulations trial moves generally consist of spin rotations about the initial direction. For micromagnetic simulations, where the magnetization length is not fixed, we must compound this by an additional trial move, namely a change in magnetization length. A Monte Carlo iteration then consists of a sequence of trial moves, exactly one per micromagnetic magnetization vector (rotation and length change), each of which is accepted with a given probability. In the Markov chain Monte Carlo approach, a sequence of ensembles is generated which converges to the required thermodynamic equilibrium distribution. A sufficient condition for convergence is that of detailed balance:

$$f(\mathbf{m}_A)W(A \to B) = f(\mathbf{m}_B)W(B \to A). \tag{2}$$

Here $W(A \to B)$ is the transition probability from state $A$ to state $B$ in the Markov chain. Thus we obtain the following ratio:

$$\frac{W(A \to B)}{W(B \to A)} = \frac{m_B^2}{m_A^2} \exp(-\Delta F / k_B T),$$

$$\Delta F = F(\mathbf{m}_B) - F(\mathbf{m}_A). \tag{3}$$

The Markov chain transition probability is given by:

$$W(A \to B) = p(A)P_{accept}(A \to B),$$

$$p(A) \propto m_A^2. \tag{4}$$

Here $P_{accept}(A \to B)$ is the trial move acceptance probability, and $p(A)$ is the conditional probability. For atomistic spin lattice simulations we simply have $p(A) = 1/N$, where $N$ is the total number of spins, i.e. each spin receives an equal weight. For micromagnetic magnetization vectors however, the weight is proportional to $m^2$, as may be seen by inspecting the partition function in Equation (1). Thus we arrive at the following acceptance probability which satisfies Equation (3), and hence detailed balance:

$$P_{accept}(A \to B) = min\left\{1, \frac{m_B^4}{m_A^4} \exp(-\Delta F / k_B T)\right\}. \tag{5}$$

In Equation (1) we now separate the longitudinal energy contribution, and rewrite it as:

$$f(\mathbf{m}) = \exp(-F(\hat{\mathbf{m}})/k_B T)f_l(m)Z_l / Z. \tag{6}$$

Here $f_l(m)$ is the magnetization length probability distribution, with $Z_l$ a renormalization factor, given by:

$$f_l(m) = \frac{m^2}{Z_l} \begin{cases} \exp\left(-\frac{VM_{S0}}{8\tilde{\chi}_\| m_e^2 k_B T}(m^2 - m_e^2)^2\right) & ,T < T_C \\ \exp\left(-\frac{VM_{S0}m^2}{2\tilde{\chi}_\| k_B T}\left(1 + \frac{3T_C m^2}{10(T-T_C)}\right)\right) & ,T > T_C \end{cases}. \quad (7)$$

The longitudinal term contribution in Equation (7), where $V$ is the micromagnetic computational cellsize volume (e.g. $V = h^3$ for a cubic cell with cellsize $h$), is the same which appears in the sLLB or LLB equation, and which gives rise to a longitudinal field. It is parametrized using the relative longitudinal susceptibility given as [5]:

$$\tilde{\chi}_\|(T) = \frac{\mu}{k_B T}\frac{B'(x)}{1 - B'(x)(3T_C/T)}, \quad x = m_e 3T_C/T. \quad (8)$$

Here $B(x) = \coth(x) - 1/x$ is the Langevin function, $\mu$ is the atomic moment, $T_C$ is the Curie temperature, and $m_e = M_e / M_{s0}$ is the normalized temperature-dependent equilibrium magnetization length given as [5]:

$$m_e(T) = B\left[m_e \frac{3T_C}{T} + \frac{\mu\mu_0 H_{ext}}{k_B T}\right]. \quad (9)$$

The remaining terms which contribute to the free energy $F(\hat{\mathbf{m}})$ in Equation (6), include all the usual micromagnetic terms, and these may be obtained directly from the corresponding energy density terms by multiplying with the computational cellsize volume $V$. There are a few terms which require special attention however, namely exchange interactions (direct and DMI) and demagnetizing interaction. First we treat the direct exchange interaction, and discuss the remaining terms later. The micromagnetic exchange free energy for a given magnetization vector $i$ is given below, where $A$ is the temperature-dependent exchange stiffness:

$$F_{ex,i} = -V \frac{2A}{M_e^2} \mathbf{M}_i . \nabla^2 \mathbf{M}_i . \tag{10}$$

It is important to note the multiplicative factor of 2 here: the micromagnetic exchange energy density (see e.g. Ref. [17]) expression is derived from the atomistic Heisenberg direct exchange Hamiltonian contribution which includes compensation for double-counting of spins. In order to calculate the free energy contribution of a single magnetization vector however, just as with the atomistic Monte Carlo method, the factor of 2 must be restored.

Single-site energy terms are obtained directly from the corresponding micromagnetic energy density expressions [17], for example the uniaxial magnetocrystalline anisotropy is included as:

$$F_{Uan,i} = V K_1 \left[ 1 - (\mathbf{m}_i . \mathbf{e}_\mathbf{A})^2 \right] + K_2 \left[ 1 - (\mathbf{m}_i . \mathbf{e}_\mathbf{A})^2 \right]^2 . \tag{11}$$

Here $\mathbf{e}_\mathbf{A}$ is the symmetry axis direction, $K_1$ and $K_2$ are the 2nd and 4th order uniaxial anisotropy constants respectively. Cubic magnetocrystalline anisotropy is included as:

$$F_{Can,i} = V K_1 \left[ \alpha_i^2 \beta_i^2 + \alpha_i^2 \gamma_i^2 + \beta_i^2 \gamma_i^2 \right] + K_2 \alpha_i^2 \beta_i^2 \gamma_i^2 . \tag{12}$$

Here $K_1$ and $K_2$ are the 4th and 6th order cubic anisotropy constants respectively, $\alpha_i$, $\beta_i$, and $\gamma_i$ are direction cosines of the magnetization vector.

Thus the MMC method consists of the following steps: 1) for a vector $i$ perform a rotation trial move in a cone of given solid angle. 2) For the same vector $i$ perform a magnetization length change trial move. 3) Compute the total free energy change, including the longitudinal free energy change using the energy term appearing in the exponent of Equation (7). 4) Accept the compound trial move with probability given in Equation (5). This procedure is done exactly once for each micromagnetic vector, which completes an MMC iteration. The algorithm is fully parallel, implemented with the red-black checkerboard decomposition scheme discussed previously for the atomistic Monte Carlo algorithm [21,22], and is thus suitable for computations on graphical processing units (GPUs). The final question is how the solid angle for the rotation trial moves should be chosen, and how the magnetization length should be changed. With time-quantized Monte Carlo (TQMC) [23,24], the cone solid angle is set such that real-time processes may be reproduced, for example the effect of field sweep rate on hysteresis loops [25], or thermally activated grain reversal time [26]. This may also be possible with MMC, however this topic is outside the scope of the current work. Instead, the cone solid angle is adaptively set such that the acceptance rate is kept at an optimal value of ~0.5 [22], which allows rapid thermalization across the entire applicable temperature range.

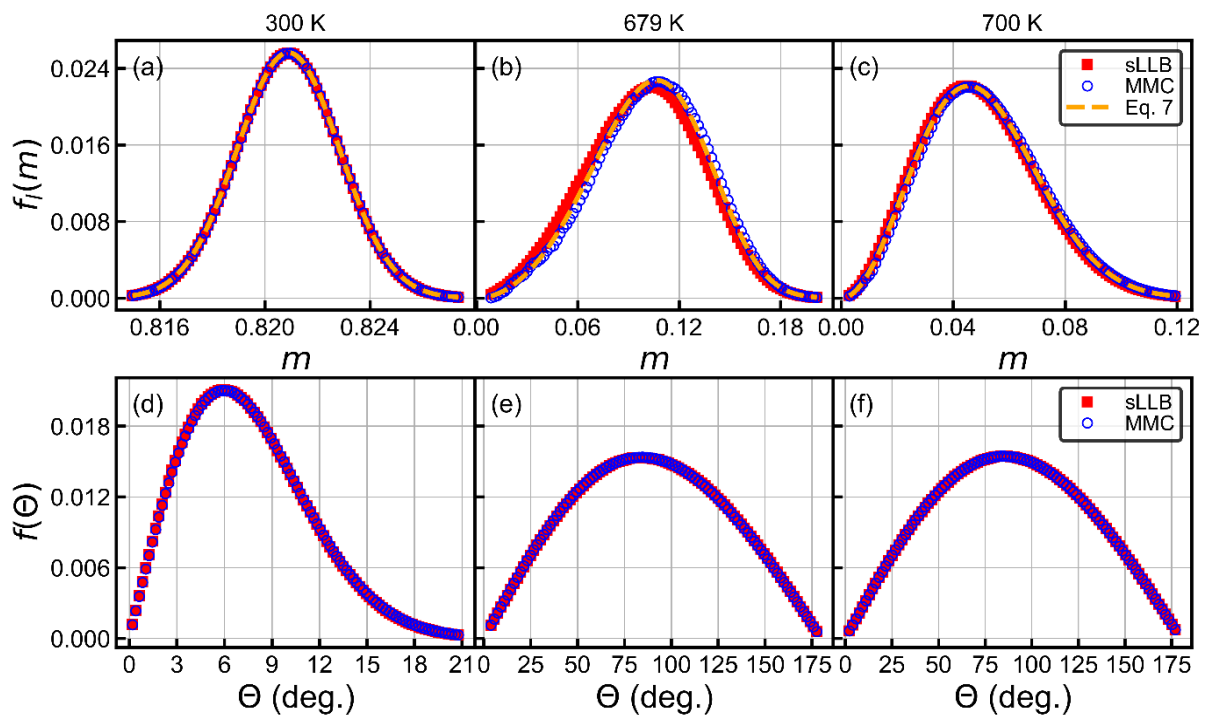

**Figure 1** – Verification of single-domain thermodynamic equilibrium properties produced by MMC, compared to sLLB. Examples are shown for 300 K (left panels), 679 K (middle panels), and 700 K (right panels), where $T_C$ = 680 K. (a)-(c) Normalized magnetization length probability distributions for indicated temperatures. (d)-(f) Histograms of angular deviation from mean direction, weighted by solid angle unit sphere differential surface area. These are equilibrium distributions averaged over many ensembles (>50,000), which do not depend on damping constant or integration time-step in the sLLB equation.

The acceptance rate is computed every 100 iterations, and the cone angle adjusted by 1° up or down if an acceptance tolerance threshold of ±0.05 is exceeded, in order to decrease or respectively increase the acceptance rate. During thermodynamic equilibrium this typically results in a cone angle variation around a mean value with a standard deviation of ~1°. The magnetization length change is performed by multiplying with a random factor uniformly chosen in the range [1 - $\sigma$, 1 + $\sigma$], where 0 < $\sigma$ < 1. At low temperatures $\sigma$ should be small in order to avoid excessive rejection of trial steps, whilst at high temperatures $\sigma$ should be large enough to allow for rapid convergence to thermodynamic equilibrium. A good choice for $\sigma$ may be obtained by noticing from Equation (7) $f_l(m)/m^2$ is a Gaussian distribution of $m^2$ with mean $m_e^2$ and width $\sigma$ given by:

$$\sigma = 2m_e \sqrt{\frac{\tilde{\chi}_\parallel k_B T}{V M_{S0}}}. \tag{13}$$

In practice this needs to be capped to a maximum value, and we use $\sigma \leq 0.03$ which is reached close to $T_C$, and also for $T > T_C$ this latter constant value is used. It should be noted that since the algorithm generates a magnetization length distribution governed by Equation (7), the probability of generating a vector with $|\mathbf{M}| > M_{S0}$ is practically zero.

We now verify the MMC method correctly reproduces the target probability distribution, by comparison with the sLLB equation. For this purpose a 50 × 50 × 50 nm simulation space is chosen with parameters typical for $Ni_{80}Fe_{20}$ ($M_{s0} = 800$ kA/m, $A(T) = A_0 m_e^2$ [27] with $A_0 = 13$ pJ/m); a value of $T_C = 680$ K was set. The direct exchange interaction is enabled with periodic boundary conditions in all directions, and a magnetic field of 10 kA/m is applied. The sLLB equation is implemented as given in Ref. [17], evaluated using the Heun method with fixed time-step of 2 fs. A zero-temperature damping value of 0.1 was set, however the equilibrium thermodynamic distributions do not depend on damping factor, nor on the integration time-step as we have verified (a small enough time-step is required for numerical convergence however, and also very small – less than 0.001 – and very large – greater than 0.5 – damping values are difficult to accommodate with good numerical precision). Results for the magnetization length distribution, averaged over >50,000 ensembles following an equal number of thermalization iterations, are shown in Figure 1(a)-(c). Very good agreement between the MMC and sLLB solutions are obtained ($R^2$ values > 0.99), also in agreement with the expected distribution in Equation (7) both below and above $T_C$. A slight discrepancy for sLLB exists close to $T_C$, reflecting difficulty in accurate numerical evaluation, however the more accurate MMC result is in excellent agreement with the theoretical distribution even around $T_C$. The equilibrium magnetization – Equation (9) – and relative longitudinal susceptibility – Equation (8) – input functions are shown in Figure 2, compared with the values obtained by fitting numerically computed distributions, as shown in Figure 1, with Equation (7). This shows the MMC method correctly reproduces the magnetization length probability distribution across the entire simulated temperature range.

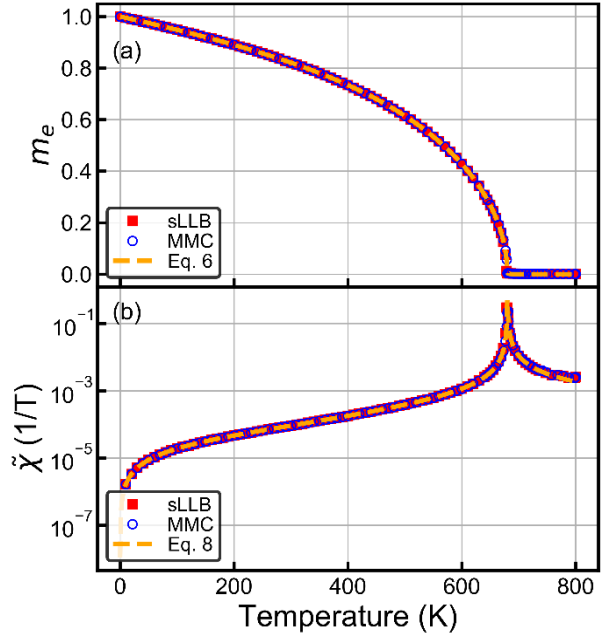

**Figure 2** – (a) Normalized equilibrium magnetization length, and (b) relative longitudinal susceptibility, computed numerically and compared to model input functions for verification. The distributions produced by MMC and sLLB respectively are fitted to obtain to obtain $m_e$ and $\tilde{\chi}$ (with Equation (7) for the longitudinal distribution).

Finally, we verify the transverse distribution in Equation (6) is reproduced correctly. For this, the angular deviation from mean direction probability distribution is computed. Thus for each micromagnetic magnetization configuration the mean direction is obtained, and for each vector in the ensemble the polar angle $\theta$ to this mean direction is found. Using 100 bins for $\theta$, between the minimum and maximum $\theta$ values, a histogram is obtained. This is then averaged over >50,000 ensembles. Example results are shown in Figure 1(d)-(f), both for MMC and sLLB, showing excellent agreement. Note, the histograms thus obtained are weighted by the bin solid angle differential surface area on the unit sphere, hence the computed probability tends to zero as $\theta$ tends to zero or to $\pi$ radians.

## III. Hysteresis Loop Modelling

One important application for MMC is finite-temperature hysteresis loop modelling. Here we concentrate on the static hysteresis limit; using kinetic Monte Carlo [28,29] it is possible to simulate the effect of field sweep rate. Alternatively TQMC may be employed, and time quantization may also be extended to MMC, however this is left for a future work. For computation of hysteresis loops, in general it is essential the demagnetizing interaction is included, allowing for shape anisotropy effects. Such long range interactions are notoriously difficult to include in a parallel Monte Carlo algorithm. Since each spin interacts with all other spins, an exact computation of the interaction energy is not compatible with domain decomposition methods. For the atomistic spin lattice case the dipole-dipole interaction was previously approximately included using a complex stream processing domain decomposition method [30]. Here we discuss a simpler alternative approach, which only requires a single demagnetizing field computation per MMC iteration, and may be achieved using the usual efficient FFT-based convolution method [17,31]. The demagnetizing field may be obtained using the discrete convolution sum shown below, where **N** is the demagnetizing tensor:

$$\mathbf{H}_{d,i} = -\sum_j \mathbf{N}_{ij}\mathbf{M}_j,$$

$$F_{d,i} = -V\mu_0 \mathbf{M}_i . \mathbf{H}_{d,i}.$$

(14)

The free energy contribution of magnetization vector $i$ is given by $F_{d,i}$, where as for the exchange interaction the value is compensated for the factor of 1/2 arising from double counting of magnetization vectors in the usual demagnetizing energy density formula. At the start of an MMC iteration the demagnetizing field is fully computed, and for each trial move the energy change $\Delta F_{d,i}$ is obtained from Equation (14) as $\Delta F_{d,i} = -V\mu_0 \Delta\mathbf{M}_i . \mathbf{H}_{d,i}$. Once a trial move is accepted, the demagnetizing field is not immediately updated. Thus whilst this approach is also inevitably an approximation, it may be made more accurate by having more than one demagnetizing field update per MMC iteration; for example since a red-black checkerboard decomposition scheme is used, the demagnetizing field can be updated before each red and black parallel passes. In turn, the red and black passes can be further decomposed to include more demagnetizing field updates. This approach was numerically tested extensively, also against a serial algorithm implementation where $\Delta F_{d,i}$ was computed exactly.

Results comparing the parallel and serial MMC implementations are shown in Appendix B, with no statistical difference found between them. This method works since locally the demagnetizing interaction is relatively weak, and the change in demagnetizing field from one iteration to another is small. Thus, in thermodynamic equilibrium, locally the trial move acceptance probability is largely driven by the exchange interaction, with $|\Delta F_{ex,i}|$ being over 1 to 2 orders of magnitude larger than $|\Delta F_{d,i}|$, even for ultrathin films with perpendicular magnetization (see Appendix B). The effect of the demagnetizing field becomes apparent only over many iterations and on a large spatial scale, appearing as a bias in the acceptance probability.

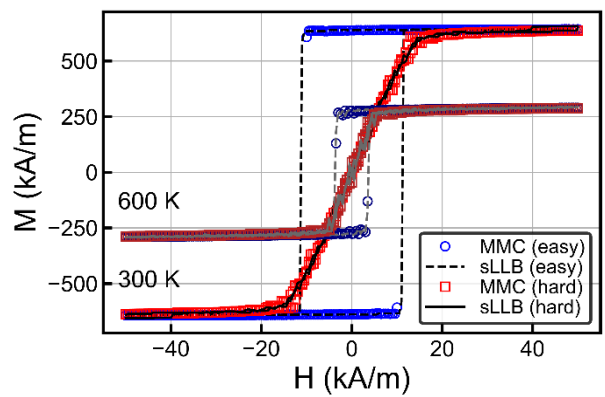

**Figure 3** – Hysteresis loops computed in a $400 \times 200 \times 5$ nm $Ni_{80}Fe_{20}$ ellipse at 300 K and 600 K, compared between the MMC and sLLB methods, along the easy axis and in-plane hard axis with 200 field steps. A shape anisotropy arises due to the demagnetizing interaction, with no magnetocrystalline anisotropy included.

First, we take a simple example to verify the shape anisotropy effect is correctly reproduced, by comparing the MMC method with the sLLB equation solution. Easy and hard axis hysteresis loops in a $400 \times 200 \times 5$ nm $Ni_{80}Fe_{20}$ ellipse, with 5 nm cubic discretization cellsize, are computed at 300 K and 600 K, with results in Figure 3 showing excellent agreement. Note, for all the hysteresis loops shown in this work only the increasing field sweep has been simulated, with the decreasing field sweep completed through inverse symmetry. For sLLB, at each field step the magnetization configuration solution is accepted when the average normalized torque value falls below $10^{-4}$ (smaller values lead to excessive simulation times). With the MMC method we need to ensure enough iterations are used to obtain correct thermodynamic equilibrium states. From atomistic Monte Carlo simulations of hysteresis loops, it is known the switching field estimation becomes increasingly more accurate as the number of Monte Carlo iterations increases [2]. One could run the MMC algorithm with a fixed number of iterations for each field step. Here a more efficient approach is taken; a chunk with a fixed number of iterations (2000) is defined, with an ensemble average magnetization value computed for each chunk. For each field step the MMC algorithm is then executed for at least 2 chunks of iterations, and if their respective final average magnetization values along the applied field direction are close enough within an acceptance threshold (0.01 normalized change acceptance threshold used), the computation accepts the last chunk average magnetization and proceeds to the next applied field value. This adaptive approach ensures a small number of iterations are used at parts of the hysteresis loop which change slowly, and a large number of iterations are expended at switching events, or at steep parts of the hysteresis curve.

As a further test we show the hysteresis loops obtained for a $400 \times 200 \times 100$ nm $Ni_{80}Fe_{20}$ ellipsoid at 300 K, also discretized using a 5 nm cubic cellsize, computed along the easy, medium, and hard axes. Here MMC is compared with results obtained using the steepest descent (SDesc) energy minimizer [3], where field steps are considered solved when the maximum normalized torque value falls below $10^{-5}$. This is used as a quasi-zero temperature method, where the material parameters are simply rescaled for the required temperature, but otherwise the SDesc method does not include stochasticity or a magnetization length distribution. A very good agreement is observed for the medium and hard axes, with hysteresis loop behaviour largely dictated by the shape anisotropy effect. A reasonable agreement is also obtained for the easy axis hysteresis loop, however here stochasticity also plays a part, resulting in the magnetization configuration switching to a vortex state close to zero field sooner for MMC compared to SDesc; the zero-field vortex state is shown in the inset to Figure 4(a). In Appendix A the easy-axis ellipsoid hysteresis loops are plotted at higher temperatures, showing the SDesc solution becomes increasingly inaccurate as the temperature increases, thus necessitating the use of MMC.

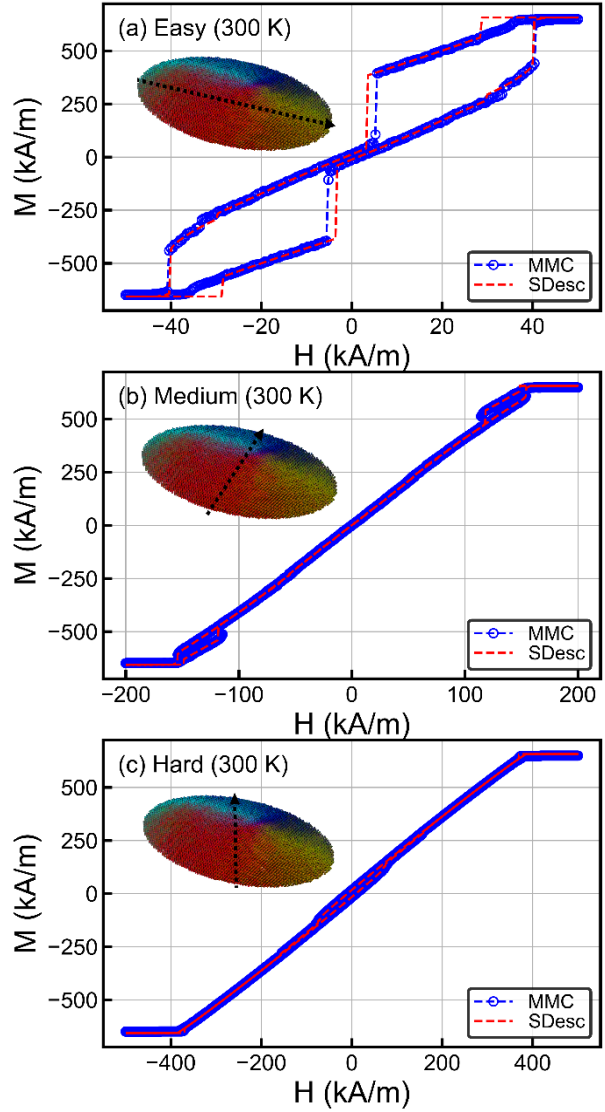

**Figure 4** – Hysteresis loops computed in a 400 × 200 × 100 nm $Ni_{80}Fe_{20}$ ellipsoid at 300 K, compared between the MMC and SDesc methods, along the indicated (a) easy, (b) medium and (c) hard axes. For (a) 250 field steps were used, and for (b), (c) 1000 field steps were used. The SDesc energy minimization solver is applied with material parameters rescaled for T = 300 K, but otherwise does not include stochasticity and the magnetization length is constrained to the equilibrium value. The insets show the vortex state obtained at zero field from MMC.

Finally we test the MMC method by computing the hysteresis loop in a large-scale granular Fe film with cubic anisotropy ($M_{s0}$ = 1.71 MA/m, $A_0$ = 21 pJ/m, $K_1$ = 48 kJ/m³ 4th order anisotropy constant, $K_2$ = -10 kJ/m³ 6th order anisotropy constant, and $T_C$ = 1044 K). For the anisotropy constants a $m_e^{l(l+1)/2}$ scaling law [32] is used, with *l* being the anisotropy term order. The simulation space consists of a 800 nm × 800 nm × 20 nm mesh with periodic boundary conditions in the plane, and cubic discretization cell with 2 nm cellsize (total of 1.6 million cells included in simulation). The grains are generated using three-dimensional Voronoi tessellation with average 10 nm spacing between seed points and with phase separation (i.e. grains interact through the demagnetizing field only). A typical granular structure generated is shown in the inset to Figure 5, with the resulting hysteresis loops obtained from MMC and temperature-rescaled SDesc also shown for $T = 300$ K. The field was applied along the in-plane horizontal direction (along cubic easy axis). A good agreement is obtained between the two methods, confirming the demagnetizing interaction is correctly evaluated in the implemented MMC method. Granular films are important for magnetic storage media, including heat-assisted magnetic recording [33]. Specialized methods exist for computing hysteresis loops in granular media, including a thermal activation model [34] and micromagnetic kinetic theory [35]. The MMC approach presented here however, is a general-purpose method for computing thermodynamic equilibrium states; the resulting collection of states are the same as those that would be obtained by integrating the sLLB equation given sufficient simulation time. For relaxation problems the MMC method is over 2 to 3 orders of magnitude faster compared to sLLB. We also note for the granular film problem of Figure 5, the MMC method is slightly faster (typical computation time ~1 h on a modern GPU) than the SDesc method, which requires many iterations to reach the normalized torque convergence threshold of $10^{-5}$.

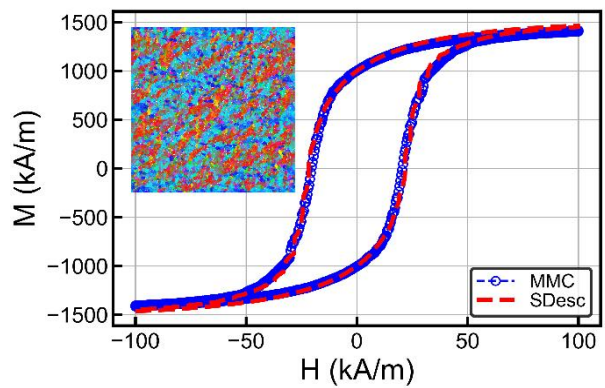

**Figure 5** – Hysteresis loop in a cubic anisotropy Fe film with 20 nm thickness, and granular structure with 10 nm average grain size. The simulation space contains over 12000 grains, discretized using a 2 nm cubic cellsize. The hysteresis loops are computed at 300 K with the MMC and SDesc methods respectively, using 500 field steps. The inset shows the magnetization configuration obtained from MMC at the coercive field (red denotes grains pointing right, and blue pointing left).

# IV. Magnetic Thin Films with DMI

The free energy contribution of the interfacial DMI term is given below, which is included in addition to the direct exchange contribution of Equation (10).

$$F_{iDMI,i} = -V \frac{2D}{M_e^2} \mathbf{M}_i \cdot \left( \hat{\mathbf{z}} \nabla \cdot \mathbf{M}_i - \nabla M_{z,i} \right) \tag{15}$$

Here we simulate the finite-temperature hysteresis loops in thin chiral Co/Pt films ($M_{s0} = 600$ kA/m, $A_0 = 10$ pJ/m, uniaxial anisotropy with perpendicular easy axis and 2$^{nd}$ order anisotropy constant $K_u = 300$ kJ/m$^3$, and $T_C = 600$ K), where the interfacial DMI is included as $D(T) = D_0 m_e^2$ [36] in Equation (15), with $D_0 = -3$ mJ/m$^2$ being the zero-temperature micromagnetic DMI constant. The simulation space is taken as 1000 nm × 1000 nm × 2 nm with periodic boundary conditions in the plane, and cubic discretization cell with 2 nm cellsize. Hysteresis loops at 300 K, 350 K, and 400 K are shown in Figure 6, where the field is applied perpendicular to the plane. The hysteresis loops are typical of experimental results [37]. A labyrinth domain structure is observed at zero field in all cases, resulting in sheared hysteresis loops. Inspecting the increasing field sweep, for $T = 300$ K the labyrinth domain structure is suddenly formed at a negative field, nucleated through thermal activation, as indicated in the inset to Figure 6(a). As the field strength is increased the labyrinth domain structure is gradually reduced, eventually forming a small number of skyrmions; further increasing the field results in thermal annihilation of skyrmions, leading to saturation. The magnetization reversal process depends on the sample temperature, not only due to increased effect of stochasticity, but more importantly due to the temperature dependence of the DMI constant and effective anisotropy. Thus at higher temperatures, instead of nucleating a labyrinth domain structure, first skyrmions are nucleated, which then grown into a labyrinth structure as shown in Figure 6(b),(c). Also at higher temperatures, increasing the field towards saturation results in a dense array of skyrmions formed out of the labyrinth structure, approximately arranged into a hexagonal lattice – see inset in Figure 6(b). The thermodynamic equilibrium states arising at each field step for the hysteresis loops in Figure 6 are detailed in the Supplemental Information.

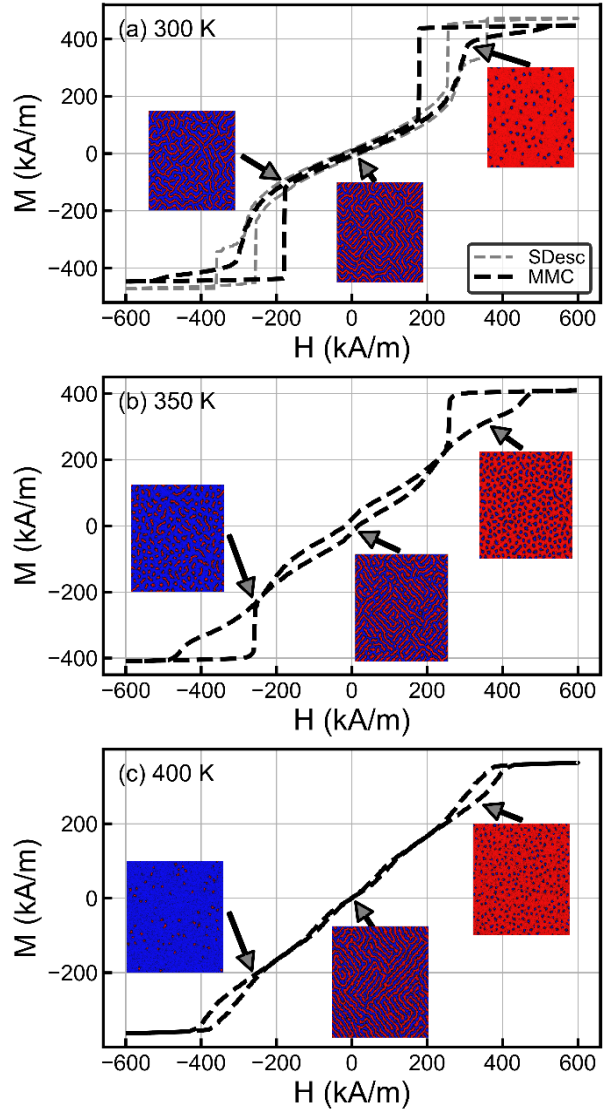

**Figure 6** – Hysteresis loops in a thin Co film with interfacial DMI and perpendicular anisotropy, $1000 \times 1000 \times 2$ nm with in-plane periodic boundary conditions, computed for (a) 300 K, (b) 350 K, and (c) 400 K, using 600 field steps. The insets show the perpendicular magnetization component at the indicated points on the increasing field sweep, with blue denoting magnetization into the plane, and red out of the plane. In (a) the SDesc energy minimizer solution is shown for comparison.

Simulation of such hysteresis loops in chiral films, with rotational symmetry about the applied field axis, is problematic with quasi-zero temperature methods such as SDesc, since the nucleation and annihilation of skyrmions and labyrinth domain structure is principally a thermally activated process. With the SDesc method, generation of a labyrinth domain structure from a uniform state purely by energy minimization requires breaking of the rotational symmetry and topological protection, and is strongly dependent on numerical floating point errors. Thus the SDesc hysteresis loop shown in Figure 6(a) is neither in quantitative nor qualitative agreement with the MMC simulations, also failing the reproduce the gradual transition from labyrinth domain structure to skyrmions, and gradual thermal annihilation of skyrmions towards saturation; instead a sudden jump is observed from a near-labyrinth domain structure to saturation.

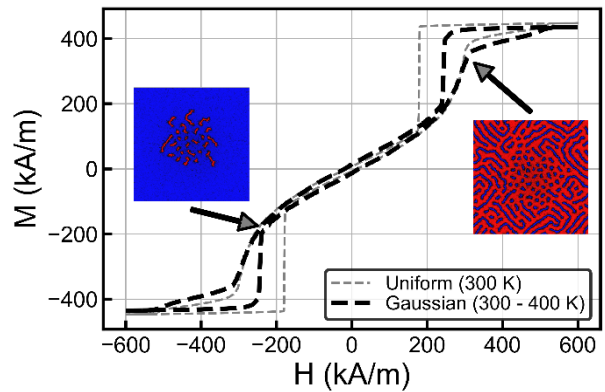

**Figure 7** – Hysteresis loop in the same Co film of Figure 6, using instead a Gaussian temperature profile with 300 nm width, reaching a maximum of 400 K at the centre, from 300 K at the extremities. The insets show the perpendicular magnetization component at the indicated points on the increasing field sweep, with blue denoting magnetization into the plane, and red out of the plane.

Finally, we also compute the hysteresis loop with a non-uniform temperature profile, in particular a Gaussian temperature profile as in Ref. [38], with 300 nm width, reaching a maximum of 400 K at the centre, from 300 K at the extremities. Such non-uniform temperature profiles are encountered in ultrafast laser-induced skyrmion nucleation studies [37-40]. Whilst the MMC method evidently does not reproduce the dynamics, it may be used to analyse the states resulting on a long time-scale, such as mixed labyrinth domain and skyrmion states. Results are shown in Figure 7. In contrast to the simulation with a uniform temperature, where a labyrinth domain structure is nucleated, instead a number of skyrmions are nucleated inside the Gaussian profile at a larger negative field, which then grow into a labyrinth domain structure. On the positive side of the hysteresis loop, first a number of skyrmions are formed from the labyrinth structure inside the Gaussian profile, resulting in a mixed state as noted in a recent experimental study [37].

The MMC method may thus be used to study thermodynamic equilibrium states also for materials with DMI, which includes hysteresis loop modelling, but also relaxed states in magnetic nanostructures, for example as observed in a recent study on Co/Pt multi-layered nano-dots [41]. For such multi-layered structures the demagnetizing field used for the MMC method may also be computed using the multi-layered convolution algorithm [31]. Analysis of skyrmion arrangements and temperature-driven diffusive motion in magnetic nano-structures was previously done using a quasi-particle Monte Carlo model [42]. The MMC method may also be used for such studies to analyse the probability distribution of skyrmion positions, having the advantage of treating the energy terms on an equal footing with micromagnetic approaches, with full spatial resolution. Thus in particular, the MMC method naturally takes into account skyrmion deformations, thermal nucleation and annihilation of skyrmions.

# V. Artificial Spin Ices

Artificial spin ices (ASI) is an active and growing field of research; for recent reviews see Refs. [43,44]. Here we briefly point out the applicability of the MMC method to the study of ASI. Monte Carlo methods have been successfully applied to the study of ASI, relying on a point dipole approximation for the magnetic moments of islands comprising ASI arrays [45,46]. On the one hand this allows computation of the blocking temperature [47], $T_B$; annealing with $T > T_B$ (normally $T_B < T_C$) allows the magnetic islands to settle into a different overall energy state, as the energy barriers for flipping their moment directions are overcome, dictated by magnetic frustration due to the dipolar interaction. Use of the dipole approximation Monte Carlo method also allows study of statistics of the vertex population types (vertex configurations grouped by energy state). The MMC method presented here may also be used, having the advantage of not relying on a dipole approximation for the magnetic islands moments. This allows for example resolving the domain configuration of magnetic islands, which may not be in a single domain state, particularly for larger magnetic islands. Effects due to magnetic islands shape can in principle be taken into account, and also connected ASI arrays may be studied, where the exchange interaction contributes to frustration. Thus the MMC approach is a more general method compared to the point dipole Monte Carlo approach; the computational complexity is greater, however efficient use may be made of GPUs which, as we show below, allows simulations of ASI arrays of dimensions approaching those used in experimental studies.

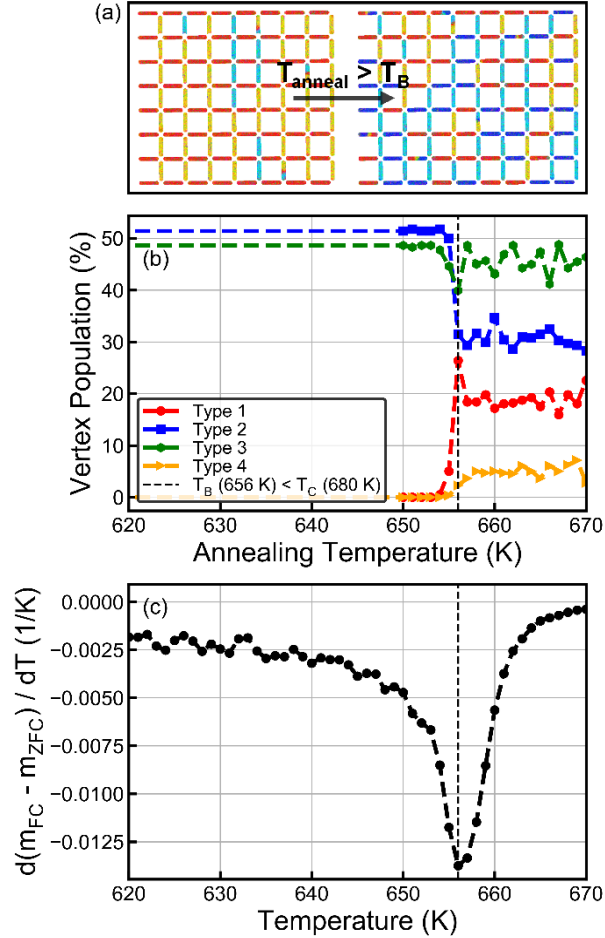

**Figure 8** – Computation of blocking temperature $T_B$, and vertex populations in square ASI. The $Ni_{80}Fe_{20}$ islands are stadium-shaped and of dimensions $200 \times 40 \times 25$ nm, with lattice spacing of 250 nm. The simulation space is 6 μm × 6 μm × 25 nm, using periodic boundary conditions in the plane and cubic cellsize of 5 nm. (a) Left-side shows a region of the simulated array before annealing, obtained at zero field after saturation. The right-side shows the same array, after annealing with a temperature above $T_B$; some elements are not in a single domain state. Red is magnetization pointing right, blue pointing left, yellow pointing up, and cyan pointing down. (b) Computation of vertex populations as a function of annealing temperature. The $T_B$ value of 656 K is indicated by the vertical dashed line. (c) Direct computation of $T_B$ using FC and ZFC curves, obtaining the same blocking temperature of 656 K as in panel (b).

In Figure 8 we show results for a square ASI array, with dimensions of 6 μm × 6 μm × 25 nm. The $Ni_{80}Fe_{20}$ islands are stadium-shaped and of dimensions $200 \times 40 \times 25$ nm, with lattice spacing of 250 nm. The simulation space was discretized using a 5 nm cubic cellsize, with in-plane periodic boundary conditions, thus resulting in 7.2 million simulation cells. An example region of the simulated ASI array is shown in Figure 8(a). Here we compute the blocking temperature $T_B$, as well as the vertex population statistics – for a square ASI, as is well known, there are 4 vertex types as given e.g. in Ref. [48]; for brevity the definitions are not repeated here. The simplest approach to computing $T_B$ is by analysing field-cooled (FC) and zero-field-cooled (ZFC) curves – see e.g. Ref. [47]. We also use this method with MMC. Thus, starting from a large temperature near $T_C$, a small field of 10 kA/m, but large enough to switch the magnetic islands along the field direction, is applied – FC – and the magnetization length in thermodynamic equilibrium along the applied field direction (horizontal direction in Figure 8) is recorded. The same is repeated in zero field – ZFC – and the temperature differential of the FC and ZFC difference curve is plotted in Figure 8(c); the point of minimum is the blocking temperature, obtained as $T_B$ = 656 K.

Next, an alternative method of computing $T_B$, which has the advantage of obtaining statistical information on vertex populations, is used. The ASI array is saturated in a large field (100 kA/m) at room temperature, along the horizontal direction, then relaxed at zero field; this starting state, showing all the horizontal islands pointing towards the right (red color) is indicated in Figure 8(a). Next, an annealing temperature is set and thermodynamic equilibrium achieved, before relaxing the ASI array at room temperature again where the vertex types are counted. This is repeated with increasing annealing temperature, and results are shown in Figure 8(b). Below $T_B$ the vertex types are exclusively Type 2 and Type 3, as these are the only possible types with the horizontal islands all pointing in the same direction. Above $T_B$ however, the magnetic moments overcome the energy barrier and settle into a different energy state, giving rise to a significant number of Type 1 vertices, largely at the cost of Type 2 vertices, although a small number of higher energy Type 4 vertices are also created. This transition is fairly abrupt at 656 K, in perfect agreement with the FC method shown in Figure 8(c). It is interesting to note, with these dimensions not all islands are in a single domain state, as seen in Figure 8(a), with a small number containing a transverse domain wall, which is stable against thermal fluctuations. Thus the MMC method may also be used to study ASI, including not only square arrays, but also kagome lattices [49] and magnetic quasicrystals [50].

# VI. Conclusions

In this work an efficient method for computing thermodynamic equilibrium states at the micromagnetic length scale was introduced, which mirrors the Metropolis Monte Carlo method commonly used for atomistic spin lattice simulations. Since micromagnetic magnetization vectors are thermodynamic averages of atomistic spins, a magnetization length distribution arises on the micromagnetic length scale. This follows a Maxwell-Boltzmann type distribution, and the micromagnetic Monte Carlo method reproduces this, in agreement with the stochastic Landau-Lifshitz-Bloch equation, by use of an additional magnetization length change trial move. Thus, the micromagnetic Monte Carlo method gives rise to the same collection of magnetization vector ensembles in thermodynamic equilibrium as the dynamic stochastic Landau-Lifshitz-Bloch equation does given sufficient simulation time. This approach is far more efficient however, allowing computation of relaxed states in large-scale systems, for which the use of the dynamic approach is not practically feasible. Particular examples have been given, including computation of hysteresis loops, both in two-dimensional and three-dimensional structures, granular films, chiral magnetic films, and study of artificial spin ices. The set of applications is not limited to these cases however, as the micromagnetic Monte Carlo method is a general approach for computing relaxed states at finite temperatures, and further work is required to fully exploit the range of applicability. Possible future extensions include use of time quantization to allow study of, for example, field sweep rate on hysteresis loops, and grain reversal times. Finally, the method introduced here was applied to ferromagnetic materials, however a future work will investigate extensions to two-sublattice models, allowing applications to ferrimagnetic and antiferromagnetic materials [51].

# Appendix A – Comparison of MMC and quasi-zero temperature solutions at high temperatures and fields

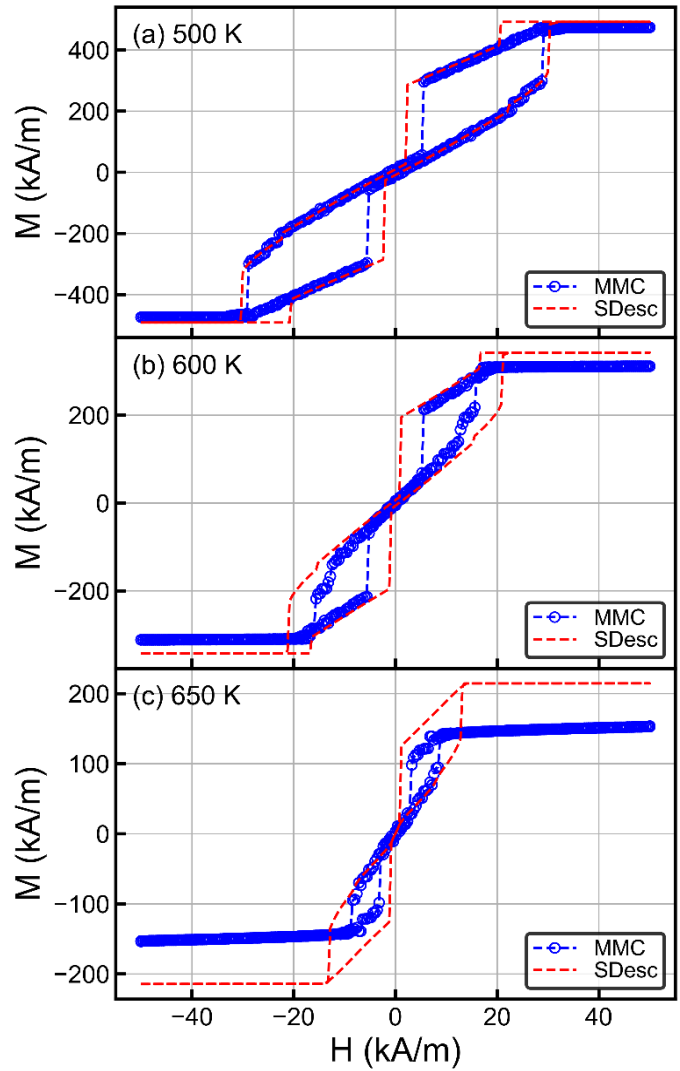

**Figure 9** – Comparison of MMC and SDesc methods for the ellipsoid problem of Figure 4, showing the easy axis hysteresis loops at higher temperatures for (a) 500 K, (b) 600 K, and (c) 650 K. As the temperature increases the SDesc solution becomes increasingly inaccurate due to lack of stochasticity and uniform magnetization length.

The SDesc and MMC methods generally agree well at low temperatures far below $T_C$, especially when magnetization reversal occurs largely through coherent rotation. An important exception to this is for processes where magnetization reversal is principally driven by thermally activated reversal, and we have given an example in the main text for chiral magnetic thin films. For such problems quasi-zero temperature methods (SDesc) are inadequate. Here we further show also for high temperatures the SDesc method is increasingly inaccurate, necessitating the use of MMC.

Figure 9 shows the $Ni_{80}Fe_{20}$ ellipsoid problem of Figure 4, with hysteresis loops given along the easy axis for 300 K, 500 K, 600 K and 650 K ($T_C$ = 680 K). As the temperature increases the MMC and SDesc solutions increasingly diverge. The difference between the solutions at high fields is due to lack of stochasticity in the SDesc method, which becomes increasingly important at higher temperatures, as the average angular deviation from the mean direction increases (see e.g. Figure 1(d)-(f)). Thus the SDesc solutions reach the magnetization saturation value at low fields, failing to reproduce a realistic high field behaviour – comparisons of high field behaviour for the SDesc and MMC solutions are shown in Figure 10. For the MMC solutions however large fields are required at high temperatures to fully saturate the magnetization along the applied field direction, requiring narrowing of the angular deviation from the mean direction probability distribution. It should be noted, for the sLLB equation form of Ref. [7] – for which the current MMC method reproduces the same thermodynamic equilibrium properties – the un-weighted average magnetization does not coincide with the equilibrium magnetization, as noted in Ref. [7] and as may easily be verified numerically. Conceptually this is problematic, since in atomistic modelling the two values are identical. This problem has been addressed recently [52], with an alternative form of the sLLB equation obtained from the LLB equation in Ref. [5]. It may be possible to extend the MMC method to this alternative sLLB formulation, however this is left for future work.

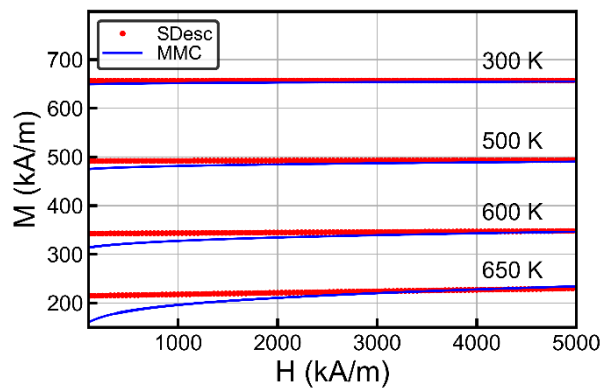

**Figure 10** – High-field behaviour of easy-axis hysteresis loops for the ellipsoid problem of Figures 4 and 9.

## Appendix B – Free energy contributions to MMC trial moves

The free energy contributions to MMC trial moves are computed a) for the Ni$_{80}$Fe$_{20}$ ellipsoid problem of Figure 4, and b) for the ultrathin Co/Pt film (here 1 nm Co thickness) of Figure 6, in order to illustrate the relative importance of the different interactions to the trial move acceptance rate. Results are shown in Figure 11 as a function of temperature, with a) the ellipsoid initialized at zero field with the vortex state shown in the insets to Figure 4, and b) the ultrathin Co film perpendicularly magnetized and with a 100 kA/m perpendicular magnetic field. The exchange interaction is the dominant contribution, with the demagnetizing interaction typically relatively negligible except when close to $T_C$. The exchange interaction term is computed using Equation (10), the demagnetizing interaction term is computed using Equation (14), and the longitudinal term is obtained from the energy term appearing in the exponent of Equation (7). This justifies the use of a single demagnetizing field update per MMC iteration, as consecutive MMC ensembles are largely correlated through the exchange interaction, with the influence of the demagnetizing interaction on thermodynamic equilibrium states being important only after many iterations (typically $10^2 – 10^4$).

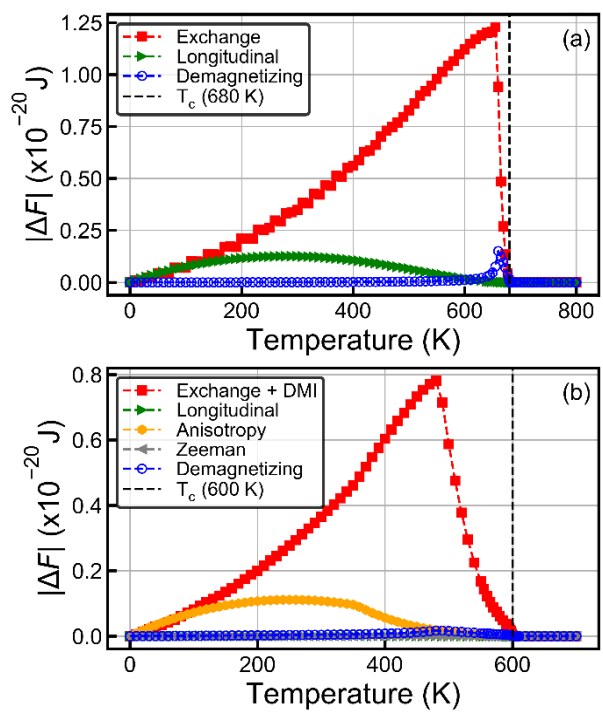

**Figure 11** – Average change in free energies for MMC trial moves, for (a) the Ni$_{80}$Fe$_{20}$ ellipsoid problem of Figure 4, initialized at zero field with the vortex state shown in the insets to Figure 4, and (b) the ultrathin Co/Pt film (here 1 nm Co thickness) of Figure 6.

Finally, we compare the parallel MMC algorithm, where the demagnetizing field is update once per iteration, with the serial MMC version where the demagnetizing field is updated after every accepted move. The latter thus exactly takes into account the long range demagnetizing interaction. Magnetization switching events are computed as a function of number of iterations for the ellipse problem of Figure 3. Starting from the remanence state with magnetization pointing along the $-x$ direction, the coercive field value of $H = 12$ kA/m is applied, and the magnetization along the field recorded over 10,000 iterations. Individual switching events are shown in Figure 12(a), as well as averages over 500 switching events for both algorithms. The switching probability is strongly influenced by the demagnetizing field, which gives rise to an effective uniaxial anisotropy. The probability of crossing the $x$ axis is plotted in Figure 12(b) as a function of iteration number. These results show there is no significant difference between the two approaches, again justifying the use of Equation (14) for including the demagnetizing interaction in the parallel MMC algorithm.

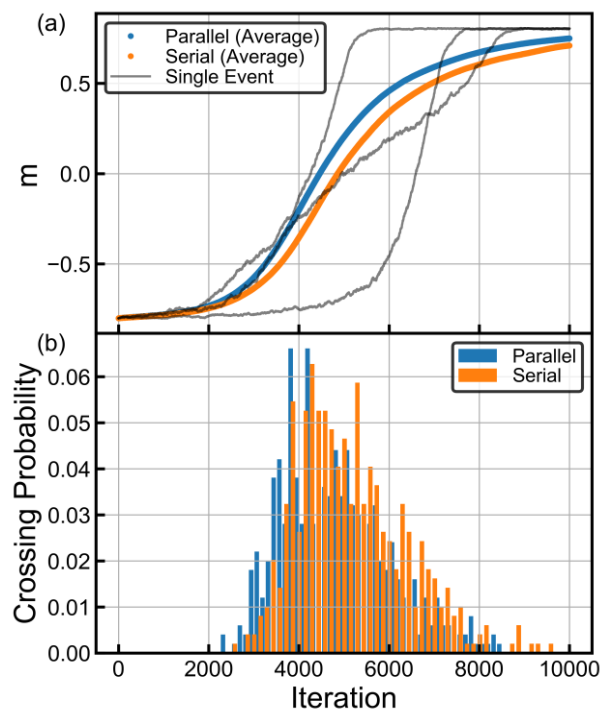

**Figure 12** – Magnetization switching for the ellipse problem around the coercive field, $H = 12$ kA/m, as a function of number of iterations. (a) Averages over 500 switching events, showing normalized magnetization as a function of iteration number, computed with the parallel MMC algorithm, as well as the serial MMC algorithm where the demagnetizing field is updated after every accepted move. Some individual switching events are also shown. (b) Probability of crossing the $x$ axis as a function of number of iterations, shown for both the parallel and serial MMC algorithms.

# Appendix C – MMC and sLLB computation times comparison

Comparisons of computation times between the MMC and sLLB methods are shown here. These are given in TABLE I for the ellipse and ellipsoid problems of Figure 3 and Figure 4 respectively. For the ellipse problem the computations were done on the central processing unit (CPU) using a Linux OS (Ubuntu) – AMD R7 3700X (16 logical cores). For the ellipsoid problem the computations were done on the GPU – Nvidia GTX 980 Ti.

For sLLB the RK4 method was used with a fixed time-step of 50 fs and a damping value $\alpha = 0.1$. For non-stochastic equations higher damping values result in faster relaxation, with an optimum damping value of $\alpha = 1.0$ due to a maximum in the damping torque (proportional to $\alpha / 1 + \alpha^2$). For sLLB however, higher damping values require smaller time-steps for integration, whilst smaller damping values result in excessive oscillations, resulting in increased computation times when relaxing; the value $\alpha = 0.1$ is a good compromise. The integration time-step was determined by computing the magnetization length probability distribution as shown in Figure 1. The value of 50 fs is the largest time-step which reproduces Equation (7), with larger time-steps resulting in significant deviations from the analytical result before any numerical divergence is observed.

For MMC the demagnetizing field is computed once per iteration as explained in the main text. For sLLB with the RK4 method, which consists of 4 sub-steps per time-step iteration, normally the demagnetizing field is computed once per sub-step. In order to improve the computation time a recently proposed method of speeding up explicit evaluation methods using demagnetizing field polynomial extrapolation [53] was used, thus also employing one demagnetizing field evaluation per iteration. As expected, computation times for MMC are significantly smaller, with speedup factors of almost 20 obtained as shown in TABLE I.

TABLE I. Comparison of computation times between MMC and sLLB for hysteresis loops of ellipse (Fig. 3) and ellipsoid (Fig. 4) problems. These are computed on the CPU, respectively GPU as indicated, with speedup factor defined as time for sLLB divided by time for MMC computation completion.

| *Problem* | MMC | sLLB | Speedup |
|---|---|---|---|
| *Ellipse (Fig. 3) – CPU* | 107 s | 1879 s | 17.6 |
| *Ellipsoid (Fig. 4) - GPU* | 747 s | 14342 s | 19.2 |

## Data Availability

The data that support the findings of this study are available from the corresponding author upon reasonable request, and the micromagnetic Monte Carlo algorithm code is available at Ref. [18].

## Supplemental Information

Videos of relaxed magnetization image sequences along the increasing field sweep are shown for each of Figure 6(a), (b), (c) and Figure 7 hysteresis loops.